\documentclass[a4paper,10pt, twoside]{article}
\usepackage[utf8]{inputenc}

\title{\LARGE A New Formulation of the Spectral Energy Budget of the Atmosphere,\\
With Application to Two High-Resolution General Circulation Models}
\author{\large\textsc{Pierre Augier,}\thanks{\textit{Corresponding author address:} 
                Pierre Augier, Linn\'e Flow Centre, KTH Mechanics, Stockholm, Sweden. 
                \newline{E-mail: pierre.augier@mech.kth.se}\vspace{2mm}
                \newline{\textit{Preprint accepted to J. Atmos. Sci. \hspace{11cm} February 15, 2013}}
                }\quad\textsc{and Erik Lindborg}\\
\textit{\footnotesize{Linn\'e Flow Centre, KTH Mechanics, Stockholm, Sweden}}
}
\date{}

\usepackage[T1]{fontenc}
\usepackage{textcomp}
\usepackage{lmodern}
\usepackage{amsmath}
\usepackage{amssymb}
\usepackage{natbib}
\usepackage{graphicx}

\usepackage[margin=12mm,top=25mm, columnsep=18pt, bottom=18mm]{geometry}
\usepackage{multicol}
\usepackage{float}

\usepackage{titlesec}
\titleformat{\section}[hang]{\bf}{\thesection}{10pt}{}
\titleformat{\subsection}[hang]{\normalfont\normalsize\itshape}{\thesubsection}{10pt}{}

\newcommand{\vv}{\textbf{v}}
\newcommand{\uu}{\textbf{u}}
\newcommand{\xx}{\textbf{x}}

\newcommand{\xxh}{\textbf{x}_h}

\newcommand{\aaa}{\textbf{a}}
\newcommand{\bb}{\textbf{b}}

\newcommand{\tuu}{\tilde\uu}
\newcommand{\tthetap}{\tilde\theta'}
\newcommand{\thetap}{\theta'}

\newcommand\p{\ensuremath{\partial}}
\newcommand{\D}{\mbox{D}}
\newcommand{\roth}{\mbox{rot}_h}
\newcommand{\divh}{\mbox{div}_h}

\newcommand{\eez}{\boldsymbol{e_z}}

\newcommand{\bnabla}{\boldsymbol{\nabla}}
\newcommand{\bnablah}{\boldsymbol{\nabla}_{\hspace{-0.6mm}h}}

\newcommand{\scalarprod}[2]{\big( #1 \, , \ #2 \big)_{lm}}

\newcommand{\scalarprodvec}[2]{\big( #1 \, , \ #2 \big)_{lm}}

\newcommand{\CC}{\mathcal{C}}
\newcommand{\GG}{\mathcal{G}}
\newcommand{\EE}{\mathcal{E}}
\newcommand{\FF}{\mathcal{F}}
\newcommand{\LL}{\mathcal{L}}
\newcommand{\DD}{\mathcal{D}}

\newcommand{\Add}[1]{#1}

\begin{document}

\maketitle

\begin{abstract}
A new formulation of the spectral energy budget of kinetic and available potential energies
of the atmosphere is derived, with spherical harmonics as base functions. 
Compared to previous formulations,
there are three main improvements: 
(i) the topography is taken into account,
(ii) the exact three-dimensional advection terms are considered 
and (iii) 
the vertical flux is separated from the energy transfer between different spherical harmonics.
Using this formulation, results from two different high resolution GCMs are analyzed:
the AFES T639L24 and the ECMWF IFS T1279L91.
The spectral fluxes show that the AFES, 
which reproduces \Add{quite} realistic horizontal spectra 
with a $k^{-5/3}$ inertial range at the mesoscales, 
simulates a strong downscale energy cascade.
In contrast, neither the $k^{-5/3}$ vertically integrated spectra nor the downscale energy cascade
are produced by the ECMWF IFS.
\end{abstract}

\begin{multicols}{2}

\section{Introduction}

The atmospheric horizontal spectra of velocity components and temperature 
show a robust $k^{-5/3}$ range at the mesoscales (10-500~km) \cite[]{NastromGage1985}. 
This power law and the corresponding one for the horizontal second order structure functions ($r^{2/3}$) 
are obtained from signals measured in the troposphere and the stratosphere, 
over both land and sea \cite[see e.g.,][]{FrehlichSharman2010}.
It still remains a challenge to reproduce these results in simulations.
Some general circulation models (GCMs) \cite[e.g.][]{KoshykHamilton2001, Hamilton_etal2008} 
and mesoscale numerical weather prediction (NWP) models \cite[]{Skamarock2004}
reproduce quite realistic mesoscale spectra.
Other GCMs, as for example ECMWF's weather prediction model Integrated Forecast System, 
produce mesoscale spectra significantly steeper 
and with smaller magnitude
than the measured ones, 
even with relatively high resolution versions \cite[e.g.,][]{Shutts2005}.
The inability of some GCMs to simulate realistic mesoscale spectra 
must have important consequences in terms of 
predictability \cite[]{VallisLIVRE2006}, 
dispersiveness of ensemble prediction systems \cite[]{Palmer2001}
and, evidently, mesoscale NWP.

Even though one can now simulate realistic mesoscale spectra, 
it is still unclear what physical mechanisms produce them.
Theoretically, the only convincing explanation of the $k^{-5/3}$ power law 
is the hypothesis that it is produced by an (upscale or downscale) energy cascade
with a constant energy flux through the scales.
Therefore, most of the different theories proposed are based on the hypothesis of an energy cascade:
upscale cascade due to 2D-stratified turbulence \cite[]{Gage1979,Lilly1983} or
downscale cascade due to internal gravity waves \cite[]{DewanGood1986, SmithFrittsVanzandt1987},
quasigeostrophic dynamics \cite[]{TungOrlando2003}, surface-quasigeostrophic dynamics \cite[]{TullochSmith2009}
or
3D-strongly stratified turbulence \cite[]{Lindborg2006}.

The principle of energy conservation strongly constrains the dynamics of the atmosphere.
In order to explain the maintenance of the general circulation, \cite{Lorenz1955} developed the concept of available potential energy (APE) and derived an approximate expression proportional to the variance of the temperature fluctuation.
His work has laid the foundations for several studies investigating APE 
\cite[e.g.,][]{Boer1989,Shepherd1993,Siegmund1994,MolemakerMcWilliams2010}
and the atmospheric energy budget through diagnostics of data from global meteorological analysis and GCMs \cite[e.g.,][]{BoerLambert2008, SteinheimerHantelBechtold2008}.
Due to the multiscale nature of atmospheric motions, 
spectral analysis can reveal valuable pieces of information 
from the data \cite[]{Fjortoft1953} and have therefore become a standard method for diagnostics.
However, drawbacks of different used formulations of the spectral energy budget severely limit the results.
First of all, the attention has been focused on the budget of kinetic energy (KE) 
so that in many studies the APE budget is not considered. 
Most studies investigate only the budget integrated over the total height of the atmosphere
and thus the vertical fluxes of energy are not computed.
Another very important limitation is that 
most studies are based on the formulation proposed by \cite{Fjortoft1953},
in which only the purely horizontal and non-divergent flow is considered
\cite[e.g.,][]{Burrows1976, BoerShepherd1983, Shepherd1987,
Boer1994, StrausDitlevsen1999, BurgessErlerShepherd2013}.
This approximation is justified for the very large scales of the atmosphere 
for which the divergence is indeed very small.
However, the approximation may lead to large errors at the mesoscales.
Atmospheric measurements show that divergent and rotational spectra are of the same order 
\cite[]{Lindborg2007jas} and
atmospheric simulations produce divergent spectra 
of the same order of magnitude as rotational spectra at the mesoscales
\cite[e.g.,][]{Hamilton_etal2008, BurgessErlerShepherd2013}.
Moreover, these results are consistent with theoretical results showing that 
strongly stratified and weakly rotating flows
tend to evolve toward states in which 
the divergent and rotational components are of the same order of magnitude 
\cite[]{BillantChomaz2001,LindborgBrethouwer2007, AugierChomazBillant2012}.
Therefore the spectral energy budget formulations based on the two-dimensional vorticity equation
can not capture the dynamics at the mesoscales.

There is no theoretical obstacle in considering the exact three-dimensional advection including both rotational and divergent components of the flow.
\cite{Lambert1984} have developed a formulation of the spectral energy budget considering both KE and APE and taking into account the exact advection term.
However, the diagnostics was integrated over the total height of the atmosphere 
so vertical fluxes were not considered.
\cite{KoshykHamilton2001} performed a diagnostic of the equation for the KE spectrum 
(the APE budget was not investigated).
The exact advection was computed but the vertical flux was not separated from the energy transfer between spherical harmonics, so it was impossible to define spectral fluxes in a conservative way.
\cite{KoshykHamilton2001} separated the spectral pressure term into adiabatic conversion and vertical flux. However, the separation was only approximate.
Moreover, as in all previous studies on the spectral energy budget, the topography was neglected.
Recently, 
\cite{BruneBecker2012} have investigated the effect of the vertical resolution in a mechanistic GCM.
Just as in \cite{KoshykHamilton2001}, 
all the terms in the kinetic spectral energy equation were computed at different pressure levels 
and it was demonstrated that they balance each other.
However, spectral fluxes were defined in a non-conservative way and actually also included vertical fluxes.

In order to investigate the energetics of the mesoscales simulated by GCMs,
it would be desirable to formulate the spectral energy budget considering both KE and APE,
taking the topography into account
and making an exact separation of the advection terms into spectral transfer and vertical flux 
and a corresponding separation of the pressure term 
into adiabatic conversion and vertical flux.
A formulation meeting these requirements is derived in section \ref{section_formulation}.
In section \ref{section_application}, results from two high resolution GCMs are analyzed.

\section{Formulation of the spectral energy budget}
\label{section_formulation}

\subsection{Governing equations in $p$-coordinates}
The analysis is performed in pressure-coordinates 
in which variables are functions of time, 
longitude $\lambda$, latitude $\varphi$
(the horizontal coordinates are denoted by $\xxh$) and pressure.
The main advantage of the $p$-coordinates is that mass conservation can be expressed 
in the same way as for an incompressible fluid: $\bnabla \cdot \vv = 0$,
where $\bnabla = (\bnablah, \p_p)$ is the gradient operator
and $\vv = (\uu,\ \omega)$ is the total velocity,
with $\bnablah$ the horizontal gradient operator, $\uu$ the horizontal velocity and $\omega = \D_t p$ the pressure velocity
($\D_t$ is the material derivative).
The hydrostatic equation is  $\p_p \Phi = - \alpha = -RT/p$,
where 
$\Phi$ is the geopotential and
$\alpha$ the volume per unit mass.

\Add{In the $p$-coordinates, the evolution equations can be written as
\begin{align}
\D_t \uu    =& - f(\varphi) \eez \wedge \uu - \bnablah \Phi + \D_\uu(\uu), \label{eq_uu}\\
\D_t H      =& \omega \alpha + \dot{Q} + \D_H(H), \label{eq_H}
\end{align}
where $f(\varphi)$ is the Coriolis parameter,
$\eez$ the upward (radial) unit vector,
$H = c_p T$ is the enthalpy per unit mass,
$\dot{Q}$ the rate of production of internal energy by heating
and 
$\D_\uu(\uu)$ and $\D_H(H)$ are diffusion terms.
The thermodynamic equation (\ref{eq_H}) can be rewritten as 
a conservation equation for} the potential temperature $\theta = \Lambda(p) T$, 
with $\Lambda(p) = (p_0/p)^\chi$, $p_0 = 1$ bar, $\chi = R/c_p \simeq 2/7$.
For simplicity, the corrections related to the vapor content are neglected.
However, the latent heat release is taken into account through the associated heating.

The main drawback of the $p$-coordinates is the complication related to the lower boundary condition 
(the topography pierces the lower pressure levels).
It can be overcome with the formalism developed by \cite{Boer1982}
(see appendix~A)
in which the dynamical equations can be written as
\begin{align}
\p_t \tuu     =& - \vv \cdot \bnabla \tuu  - f(\varphi) \eez \wedge \tuu - \beta\bnablah \Phi 
+ \beta \D_\uu(\uu), \label{eq_tuu}\\
\p_t \tthetap =& - \vv \cdot \bnabla \tthetap -  \tilde\omega \p_p \langle \theta\rangle_r  %
                + \tilde{Q}_\theta - \beta \p_t \langle \theta \rangle_r + \beta\D_\theta(\theta),
\label{eq_tthetap}
\end{align}
where 
$\D_\theta(\theta)$ is a diffusion term,
$Q_\theta \equiv \Lambda(p) \dot{Q} /c_p$
and 
$\tuu = \beta(\xxh, p) \uu$, 
with $\beta(\xxh, p)$ equal to one above the surface and to zero below.
The potential temperature fluctuation $\tthetap$ is defined as
$\tthetap = \tilde \theta - \beta \langle \theta \rangle_r$,
where $\langle \theta \rangle_r = \langle \beta \theta \rangle / \langle \beta\rangle$ 
is the representative mean, 
i.e., the mean over regions above the surface
(the brackets $\langle \cdot \rangle $ denote the mean over a pressure level).

\subsection{Kinetic and available potential energy forms}

\cite{Lorenz1955} showed that the sum of the globally integrated 
kinetic energy (KE) and available potential energy (APE) is approximately conserved.
The mean energies per unit mass are $ E_K(p) = \langle|\tuu|^2\rangle/2$ 
and $ E_A(p) = \gamma(p) \langle  \tthetap^2  \rangle/2 $, 
with $\gamma(p) = R/(-\Lambda(p)p\p_p \langle \theta\rangle_r )$.
The energy budget can be written as
\begin{eqnarray}
\p_t E_K(p) &=& \hspace{11.8mm}   C(p) + \p_p F_{K \uparrow}(p) - D_K(p) \nonumber \\
            & & + S(p), \label{eq_EKp}\\
\p_t E_A(p) &=&    G(p)        -  C(p) + \p_p F_{A \uparrow}(p) - D_A(p) \nonumber \\
            & & + J(p), \label{eq_EAp}
\end{eqnarray}
where $G(p) = \gamma(p)\langle\tthetap \tilde{Q}_\theta' \rangle$ is forcing by heating 
(differential heating at very large scales and latent heat release),
$C(p) = -\langle \tilde \omega \tilde\alpha\rangle$ is conversion of APE to KE,
$F_{A \uparrow}(p) = -\gamma(p) \langle \omega \tthetap^2 \rangle/2 $ and
$F_{K \uparrow}(p) = -          \langle \omega |\tuu|^2   \rangle/2
                     -\langle  \tilde\omega \tilde\Phi  \rangle$
are vertical fluxes,
$D_K(p)$ and $D_A(p)$ are diffusion terms.
\Add{
The last terms of (\ref{eq_EKp}) and (\ref{eq_EAp}),
$S(p) = -\langle \delta \p_t (p_s \Phi_s)  \rangle$ 
and 
$J(p) = (\p_p \gamma) \langle \omega \tthetap^2 \rangle /2 
- \langle \beta^2 \rangle  \langle\omega\rangle_r \langle\alpha\rangle_r $,
correspond to adiabatic processes which do not conserve the sum of the KE and the Lorenz APE.
However,
when integrated over the whole atmosphere, these terms are negligible so that 
the sum of the total kinetic energy and the total Lorenz APE is approximately conserved.}
\cite{Siegmund1994} showed that the globally integrated exact APE and Lorenz APE differ by less than 3\%.
\Add{
In appendix~B, we show how equations (\ref{eq_EKp}-\ref{eq_EAp}) 
can be related to the equation of total energy conservation.}%

\subsection{Spectral analysis based on spherical harmonic transform}

\Add{For clarity, the spectral energy budget is here derived for levels above the topography 
for which $\beta(\xxh, p) = 1$. 
In this section, we do not include the non-conservative term corresponding to $J(p)$,
i.e.\ we considered $\gamma$ as a constant.
The general formula used for the numerical computations are given in appendix~A.}
Each scalar function defined on the sphere can be expanded as a sum of 
spherical harmonics functions $Y_{lm}(\xx_h)$, 
which are the normalized eigenfunctions of the horizontal Laplacian operator on the sphere: 
$|\bnablah|^2 Y_{lm}= -l(l+1) Y_{lm}/a^2$
and 
$\langle Y_{l'm'}^* Y_{lm} \rangle = \delta_{ll'} \delta_{mm'}$, 
where the star denotes the complex conjugate 
and $l$ and $m$ are the total and zonal wavenumbers
\cite[for details, see][]{Boer1983}. 
\Add{The potential temperature fluctuation is written as
\begin{equation}
\thetap(\xx_h, p) = 
\sum_{l\geqslant 0} \sum_{-l\leqslant m \leqslant l} 
\thetap_{lm}(p) Y_{lm}(\xx_h),
\label{eq_decompoSH}
\end{equation}
and the other scalar variables are written in the corresponding way.
It follows that the mean over a pressure level of the product of two functions
can be written as
\begin{equation}
\langle \omega \Phi \rangle = 
\sum_{l \geqslant 0} \sum_{-l\leqslant m \leqslant l}
\scalarprod{\omega}{\Phi},
\end{equation}
where, by definition,
\begin{equation}
\scalarprod{\omega}{\Phi} \equiv 
\Re\{ \omega_{lm}^* \Phi_{lm} \},
\label{eq_def_scalarprod}
\end{equation}
$\Re$ denoting the real part.
The spectral APE function can thus be defined as
\begin{equation}
E_A^{[lm]}(p) = \gamma(p) \frac{\scalarprod{\thetap}{\thetap}}{2} = \gamma(p) \frac{|\thetap_{lm}(p)|^2}{2}
\label{eq_def_EAlm}
\end{equation}
so as the mean APE at a pressure surface is given by
$E_A(p) = \sum_{l,m} E_A^{[lm]}(p)$.
}

\Add{
The meridional and azimuthal components of a vector field on the sphere 
are multiple-valued at the poles because of the coordinate singularity.
In order to expand a vector field in spherical harmonics, 
it is thus appropriate to decompose it in terms of two scalar functions.
For example, the velocity field is decomposed as 
$\uu = -\bnablah \wedge (\psi\eez) + \bnablah \chi$,
where 
$\psi(\xx_h,p)$ is the spherical stream function 
and 
$\chi(\xx_h,p)$ the spherical velocity potential.
The vertical component of the vorticity is given by
$\zeta \equiv \roth (\uu) \equiv \eez \cdot (\bnablah \wedge \uu) = \bnablah^2 \psi$
and the horizontal divergence by 
$d \equiv \divh(\uu) \equiv \bnablah \cdot \uu = \bnablah^2 \chi$.
%
% % % % Note that since the meridional and azimuthal components are properly defined 
% % % % nearly everywhere on the sphere 
% % % % (the surface associated with the poles is zero),
% % % % it is also possible to compute the spherical harmonic coefficients of the components 
% % % % using an adapted (e.g., Gaussian) grid.
%
Our formulation is based on a result obtained from the rotational-divergent split.
The mean value over a sphere with radius $a$ (in our case the radius of the Earth) 
of the scalar product between two horizontal vector fields $\aaa$ and $\bb$
can be written as
\begin{equation}
\langle \aaa\cdot\bb \rangle = \sum_{l \geqslant 1} \sum_{-l\leqslant m \leqslant l} 
\scalarprodvec{\aaa}{\bb},
\label{eq_scalarprodvec}
\end{equation}
where, by definition,
\begin{align}
\scalarprodvec{\aaa}{\bb}
\equiv \frac{a^2}{l(l+1)} 
\Re\{ 
& \ \roth(\aaa)_{lm}^*  \roth(\bb)_{lm} \nonumber \\
 +  & \ \divh(\aaa)_{lm}^*  \divh(\bb)_{lm}
\},
\label{eq_def_scalarprodvec}
\end{align} 
which rests on the fact that $Y_{lm}$ are eigenfunctions of the Laplace operator.
Here, we have used a similar notation on the LHS of 
(\ref{eq_def_scalarprod}) and (\ref{eq_def_scalarprodvec}). 
It should be understood that 
when the two arguments of $\scalarprod{.}{.}$ are scalars, we use (\ref{eq_def_scalarprod})
and
when the arguments are two vector fields, we use (\ref{eq_def_scalarprodvec}).
From (\ref{eq_scalarprodvec}), it is clear that the spectral KE function can be defined as
\begin{equation}
E_K^{[lm]}(p) = \frac{\scalarprodvec{\uu}{\uu}}{2} = \frac{a^2 (|\zeta_{lm}|^2 + |d_{lm}|^2)}{2l(l+1)}.
\label{eq_def_EKlm}
\end{equation}}%

% % % % The spectral APE budget (\ref{eq_EAlmp}) is derived by multiplying equation (\ref{eq_tthetap}) 
% % % % by $\gamma(p) \thetap^*_{lm}$ 
% % % % and then taking the real part of the resulting equation. 
% % % 
% % % % The spectral KE budget (\ref{eq_EKlmp}) is derived by 
% % % % substituting (\ref{eq_tuu}) into the equality
% % % % $\p_t E_K^{[lm]}(p) = \scalarprodvec{\uu}{\p_t \uu} $.

\Add{
The spectral energy budget is derived by 
substituting (\ref{eq_tuu}) and (\ref{eq_tthetap})
into the time differentiations of
equations (\ref{eq_def_EKlm}) and (\ref{eq_def_EAlm}), 
that is
$\p_t E_K^{[lm]}(p) = \scalarprodvec{\uu}{\p_t \uu} $
and 
$\p_t E_A^{[lm]}(p) =\gamma(p) \scalarprod{\thetap}{\p_t \thetap} $, respectively.
Reorganizing the different terms, the spectral energy budget can be written as
\begin{eqnarray}
    \partial_t E_K^{[lm]}(p)%
 & = & \hspace{16.9mm}  C^{[lm]}(p) + T_K^{[lm]}(p)  + L^{[lm]}(p)  \nonumber \\
    &&  + \p_p F_{K \uparrow}^{[lm]}(p) - D_K^{[lm]}(p),  \label{eq_EKlmp} \\
    \partial_t E_A^{[lm]}(p) %
 & = &  G^{[lm]}(p)   - C^{[lm]}(p)  + T_A^{[lm]}(p)  \nonumber\\
    &&  + \p_p F_{A \uparrow}^{[lm]}(p) - D_A^{[lm]}(p), \label{eq_EAlmp}
\end{eqnarray}
where $T_K^{[lm]}(p)$ and $T_A^{[lm]}(p)$ are spectral transfer terms due to nonlinear interactions, 
$L^{[lm]}(p)$ is a spectral transfer term arising from the Coriolis term. 
Each of the other terms corresponds to a term in (\ref{eq_EKp}-\ref{eq_EAp}).
}

\Add{
We first focus on the nonlinear term
$ -\gamma \scalarprod{\thetap}{\vv \cdot \bnabla \thetap} $.
From the expression of $F_{A \uparrow}(p)$ in equation (\ref{eq_EAp}), 
it is clear that the APE vertical flux can be written as
\begin{equation}
F_{A \uparrow}^{[lm]}(p) =  - \gamma  \scalarprod{\thetap}{\omega \thetap}/2.
\end{equation}
A simple method to obtain the spectral transfer term 
is to compute the complementary part of the nonlinear term
\begin{equation}
T_A^{[lm]}(p)     =   -\gamma \scalarprod{\thetap}{\vv \cdot \bnabla \thetap}
+\gamma \p_p \scalarprod{\thetap}{\omega \thetap}/2.
\end{equation}
It is straightforward to show that 
the sum over all spherical harmonics of $T_A^{[lm]}(p)$ 
is equal to
$-\gamma\langle \bnablah \cdot ( \uu |\thetap|^2/2) \rangle = 0 $,
meaning that 
this transfer term is exactly conservative and only redistributes energy 
among the different spherical harmonics at a pressure level.
The diabatic term, the conversion term and the APE diffusion term are 
\begin{align}
G^{[lm]}(p)   = & \ \gamma \scalarprod{\thetap}{Q_{\theta}'},\\
C^{[lm]}(p)   = & -       \scalarprod{\omega}{\alpha}, \\
D_A^{[lm]}(p) = & -\gamma \scalarprod{\thetap}{\D_\theta(\theta)}.
\end{align}
}

\Add{
Using the continuity equation, the hydrostatic equation
and the fact that the spherical harmonics are eigenfunctions of the Laplace operator,
the pressure term is separated into conversion and vertical flux\footnote{
\cite{KoshykHamilton2001} made a similar but approximate decomposition 
using the spherical harmonics transforms of the components of the vectors.}
\begin{equation}
- \scalarprodvec{\uu}{\bnablah \Phi}  = C^{[lm]}(p) - \p_p \scalarprod{\omega}{\Phi}.
\end{equation}
The total KE vertical flux is the sum of the pressure flux plus the turbulent KE flux
\begin{equation}
F_{K \uparrow}^{[lm]}(p) = -\scalarprod{\omega}{\Phi}
- \scalarprodvec{\uu}{\omega \uu}/2.
\end{equation}
The nonlinear KE spectral transfer is computed as
the complementary part of the nonlinear terms
\begin{equation}
T_K^{[lm]}(p) = -\scalarprodvec{\uu}{\vv\cdot\bnabla \uu} 
+ \p_p \scalarprodvec{\uu}{\omega \uu}/2,
\end{equation}
which assures that it conserves energy at a pressure level.
The horizontal advection of the horizontal velocity is computed using the relation 
$\uu\cdot\bnablah \uu = \bnablah |\uu|^2/2  + \zeta \eez \wedge \uu$.
A transfer term arises from the Coriolis term
\begin{align}
L^{[lm]}(p) & =  -\scalarprodvec{\uu}{f(\varphi) \eez \wedge \uu} \nonumber\\
            & = \              \scalarprod{\psi}{\roth(f(\varphi) \eez \wedge \uu)} \nonumber\\
            &  \hspace{4mm}  +\scalarprod{\chi}{\divh(f(\varphi) \eez \wedge \uu)} . 
\label{eq_Llmp} 
\end{align}
At the $f$-plane, where $f$ is constant and a Fourier decomposition is used,
the corresponding term is zero.
Using the definition of the Coriolis parameter $f(\varphi) = f_0 \sin\varphi$
and splitting the velocity in rotational and divergent parts, one can show that 
\begin{align}
L^{[lm]}(p) = f_0 \big( &    \scalarprod{\psi}{\sin\varphi  d + \cos\varphi \p_\varphi \chi /a^2} \nonumber\\
            & \hspace{0mm} - \scalarprod{\chi}{\sin\varphi  \zeta + \cos\varphi \p_\varphi \psi /a^2}  \big).
\label{eq_Llmp_2} 
\end{align}
This implies that the linear transfer does not involve rotational-rotational (nor divergence-divergence) interactions.
}

\Add{To sum up,} in contrast to the previous formulations, here,
the APE budget is included,
the exact three-dimensional advection is considered
and 
the vertical flux terms and the horizontal spectral transfer terms are exactly separated.
\Add{Moreover, the topography, which has been neglected in this section for clarity, 
can consistently be taken into account in the spectral analysis, as shown in appendix~A.}

\subsection{Vertical integration, summation over zonal wavenumbers and cumulative summation over total wavenumbers}

Since the density strongly varies with height in the atmosphere 
we prefer to include the density when we integrate spectral energies over layers, 
so that our quantities have the dimension of energy 
rather than energy per unit mass as in most other studies.
With the formulation by Boer, this can be done easily even for pressure levels pierced by the topography.
The vertically integrated KE spectrum is defined as 
\begin{equation}
E_K[l]_{p_t}^{p_b} = \int_{p_t}^{p_b} \frac{dp}{g} \sum_{-l\leqslant m \leqslant l} E_K^{[lm]}(p).
\end{equation}
The vertically integrated nonlinear spectral flux of kinetic energy is defined as
\begin{equation}
\Pi_K[l]_{p_t}^{p_b} = \sum_{n \geqslant l} \int_{p_t}^{p_b} \frac{dp}{g} \sum_{-n\leqslant m \leqslant n} T_K^{[nm]}(p),
\label{eq_PiK}
\end{equation}
and the spectral flux of APE is defined in the corresponding way.
When (\ref{eq_EKlmp}) and (\ref{eq_EAlmp}) are vertically integrated, divided by $g$ and
summed as in (\ref{eq_PiK}) over all the spherical harmonics with total wavenumber $\geqslant l$, 
we obtain
\begin{eqnarray}
\p_t \EE_K[l]_{p_t}^{p_b} &=& %
\hspace{12.9mm}          \CC[l]_{p_t}^{p_b} +\Pi_K[l]_{p_t}^{p_b} +\LL[l]_{p_t}^{p_b}   \nonumber \\
                          & & %
+\FF_{K \uparrow}[l](p_b) - \FF_{K \uparrow}[l](p_t) -\DD_K[l]_{p_t}^{p_b} , \label{eq_flux_K} \\
\p_t \EE_A[l]_{p_t}^{p_b} &=& %
\GG[l]_{p_t}^{p_b}     -\CC[l]_{p_t}^{p_b} +\Pi_A[l]_{p_t}^{p_b}  \nonumber \\
                          & & %
+\FF_{A \uparrow}[l](p_b) - \FF_{A \uparrow}[l](p_t) -\DD_A[l]_{p_t}^{p_b} , \label{eq_flux_A}
\end{eqnarray}
where the terms named $\FF_{\uparrow}[l](p)=\sum_{n\geqslant l}  F_{\uparrow}[n]$
are cumulative vertical fluxes and
each of the other terms is an integrated cumulation of a corresponding term in (\ref{eq_EKlmp}) and (\ref{eq_EAlmp}). 
For example,
$\EE_K[l]_{p_t}^{p_b} = \sum_{n\geqslant l}  E_K[n]_{p_t}^{p_b}$ 
is the cumulative kinetic energy
and 
$\CC_K[l]_{p_t}^{p_b} = \sum_{n\geqslant l}  C[n]_{p_t}^{p_b}$ the cumulative conversion of APE into KE.
Putting $l=0$ in equations (\ref{eq_flux_K}) and (\ref{eq_flux_A}) 
we recover equations (\ref{eq_EKp}) and (\ref{eq_EAp}) integrated between two pressure surfaces.

\section{Application to two GCMs}
\label{section_application}

\subsection{Presentation of the data and the models}

We have analyzed two data sets produced 
% % % % by two simulations performed 
with two spectral GCMs:
the Atmospheric GCM for the Earth Simulator (AFES) 
and ECMWF's weather prediction model Integrated Forecast System (IFS).
For a precise descriptions of both simulations, 
see \cite{Hamilton_etal2008} and \cite{ECMWF2010}, respectively.
The two simulations and the two models are quite different.
The AFES is a climate model using a spectral advection scheme.
It has been run at high resolution for research purposes.
The horizontal resolution is T639, which corresponds to a 
\Add{minimum wavelength of roughly 60~km}.
% % % % grid space of roughly 20~km.
\Add{The model is formulated in sigma coordinates
with} 24 vertical levels from the ground to about 1 hPa
\Add{leading to a vertical grid space corresponding to approximately 2~km in the high troposphere.}
\cite{TakahashiHamiltonOhfuchi2006} and \cite{Hamilton_etal2008} showed that the AFES reproduces many features of the atmospheric spectra, 
especially a realistic $k^{-5/3}$ power law at the mesoscales. 
\Add{
For each truncation wavenumber $l_T$, 
the value of the horizontal hyperdiffusion coefficient $\kappa_h$ was adjusted by trial-and-error to produce power law
spectra that agree with observations.
However, for the runs at sufficiently large resolution (T639 and T1279) 
the existence of a distinct $k^{-5/3}$ mesoscale range was shown to be independent of the hyperdiffusion employed.
Moreover, \cite{TakahashiHamiltonOhfuchi2006} showed that the tuned horizontal hyperdiffusion coefficient 
scales with the truncation wavenumber as $\kappa_h \propto l_T^{-3.22} $.
According to energy cascade phenomenology this coefficient should only depend
on the energy flux $\Pi$, through the inertial range, and the resolution scale
so that the dissipation will take place at the smallest resolved scales
and will be approximately equal to $\Pi$. 
Dimensional considerations then give 
$\kappa_h \simeq \Pi^{1/3} l_T^{-10/3}$.
This good agreement between the theoretical and the empirically determined scaling laws 
indicates that the $k^{-5/3}$ spectra in the AFES models are produced by a downscale energy cascade
\cite[]{Hamilton_etal2008}.}
ECMWF IFS is a model developed and used for operational deterministic forecast.
It uses a semi-Lagrangian advection scheme with a horizontal resolution T1279,
which corresponds to a 
\Add{minimum wavelength of roughly 30~km}.
% % % % grid space of roughly 10~km.
\Add{The model is formulated in hybrid coordinates
with} 91 vertical levels and a minimum pressure of $1$ Pa.
\Add{This leads to a vertical grid space corresponding to approximately 500~m in the high troposphere.}
\Add{A semi-implicit time scheme makes it possible to run this model 
with a much larger time step (600 s) 
than the one used for the AFES T639 model (100 s).}
The AFES and the ECMWF simulations correspond to June and December, respectively.
We have averaged
over 10 days (40 times) for the AFES data set
and
over 25 days (5 different times) for the ECMWF data set.
\Add{
The time variations of the spectra and of the terms involved in the spectral energy budget 
are not very large, such that averages over such limited statistics represent the 
important features of the models, especially at the mesoscales 
for which a statistical convergence seems to be achieved.
Other computations for the ECMWF model for August have shown that seasonality 
can not explain the main differences between the models.}

The AFES data
\Add{are already linearly interpolated from the model levels
and consist of} %
% % %  contain 
the horizontal velocity, the pressure velocity and the temperature at pressure levels.
The geopotential is computed by integrating the hydrostatic relation from the ground.
The ECMWF data
\Add{are raw data from the model outputs. 
They} contain the vorticity, the divergence, the pressure velocity and the temperature at hybrid vertical levels.
We compute horizontal velocity and the geopotential and then linearly interpolate the data at pressure levels.
\Add{The interpolation method is thus the same for both models.}
Since we do not have access to the heating rate and to the total dissipative terms, 
the APE forcing $\GG[l]_{p_t}^{p_b}$ and the dissipative terms 
$\DD_K[l]_{p_t}^{p_b}$ and $\DD_A[l]_{p_t}^{p_b}$ have not been computed.
Since the surface fluxes are modeled in a GCM, 
we focus on the vertical flux at pressure levels not pierced by the topography.
Finally, we have computed the terms 
$\CC[l]_{p_t}^{p_b}$, $\Pi_K[l]_{p_t}^{p_b}$, $\Pi_A[l]_{p_t}^{p_b}$ and $\LL[l]_{p_t}^{p_b}$ 
for all levels
and the vertical fluxes $\FF_{K \uparrow}[l](p)$ and $\FF_{A \uparrow}[l](p)$ 
for pressure levels not pierced by the topography.
\Add{The spectral flux arising from the Coriolis term $\LL[l]_{p_t}^{p_b}$ 
is completely negligible at wavenumbers $l \gtrsim 10$,
consistent with previous computations by \cite{KoshykHamilton2001}.
At larger scales, it is not negligible and is quite similar for both models.
The Coriolis strength leads to a positive vertically integrated spectral flux 
of the order of 1~W/m$^2$
from $l\simeq 2$ to $l\simeq 6$
with dominant contribution from the stratosphere.
The linear flux $\LL[l](p)$ is small in the troposphere for all wavenumbers. 
}%
Since our main interest here is the dynamics of the mesoscales and for clarity,
this spectral flux is not included in the figures.

\subsection{Vertically integrated spectral energy budget}

\label{subsection_vert_int_SEB}

\begin{figure*}
\centering
% \centerline{
% \includegraphics[width=22pc]{Fig/fig_dyn_atm_AFES_T639_withrot.eps}
% \includegraphics[width=22pc]{Fig/fig_dyn_atm_ECMWF_T1279winter.eps}
% }
\centerline{\includegraphics[width=42pc]{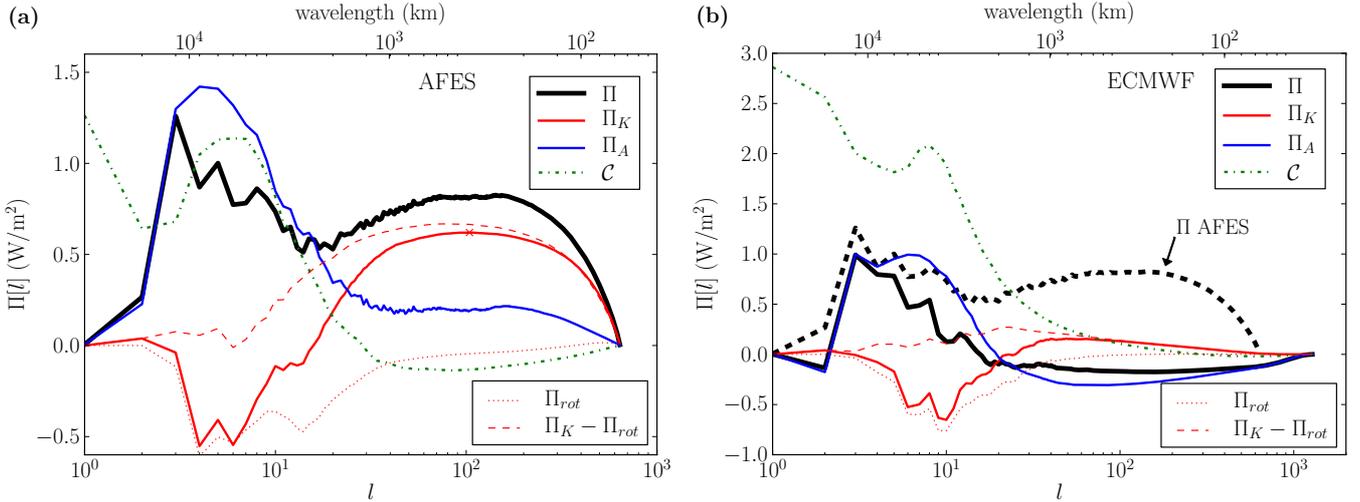}}
\caption{
Total, KE and APE nonlinear spectral fluxes 
and cumulative conversion $\CC[l]$ versus total wavenumber
for 
(a) the AFES T639 simulation and 
(b) the ECMWF IFS T1279 simulation.
The cross in (a) marks the maximum value of the KE nonlinear spectral flux 
$\tilde\Pi_K = \max({\Pi_K}) = 0.62$ used for nondimensionalization of the spectra
in figure~\ref{fig_spectra_whole_atm}.
In (b), the black dashed line is the spectral flux $\Pi[l]$ for the AFES T639 simulation.
}
\label{fig_dyn_whole_atm}
\end{figure*}

Figure \ref{fig_dyn_whole_atm}(a) presents the globally integrated spectral fluxes and cumulative conversion
for the AFES model. 
\Add{When vertically integrated over the total height of the atmosphere 
these quantities are simply denoted $\Pi_K[l]$, $\Pi_A[l]$ and $\CC[l]$,
i.e., without the top and bottom pressures in subscript and exponent.}
By construction the fluxes are equal to zero at $l=l_{\max}$ 
and should also be equal to zero at $l=0$, since they represent conservative processes. 
The total spectral flux (black thick line) is positive at all wavenumbers, 
which means that in average, the energy is transferred toward large wavenumbers.
At leading order, energy is forced at the very large planetary scales and dissipated at smaller scales, 
and a substantial part is dissipated at the smallest scales simulated.

However, the total spectral flux has a somewhat intricate shape.
It reaches a maximum equal to 1.3~W/m$^2$ around $l=4$, 
decreases to 0.55~W/m$^2$ at $l\simeq 20$, 
increases again to reach a plateau between $l\simeq 70$ and $l\simeq 200$ where $ \Pi[l] \simeq 0.82$~W/m$^2$ 
before dropping down to zero at the largest wavenumbers.
At $l<15$, the total flux is largely dominated by the APE spectral flux (blue line), 
which also increases abruptly around $l=2$ and decreases abruptly between $l=6$ and $l=20$.
This indicates that there is a transfer of APE from wavelengths of the order of 10000~km 
to wavelengths between 2000~km and 5000~km.
The strong decrease of $\Pi_A[l]$ is associated with 
a strong increase of the KE spectral flux (red line)
and a strong decrease of the cumulative conversion (dashed dotted line) 
from $\CC[l] = 1.13$~W/m$^2$ to $\CC[l] = -0.14$~W/m$^2$.
\Add{
The amount of energy which is converted from APE to KE 
in the wavenumber range $[l_1,\ l_2]$ 
is equal to $\Delta\CC \equiv \CC[l_1]-\CC[l_2]$.
(Note here the non-standard definition of the difference operator $\Delta$.)
Therefore, the strong decrease of the cumulative conversion 
over the range of wavelengths between 1000~km to 5000~km
corresponds to a conversion of APE to KE of $\Delta\CC \simeq 1.27$~W/m$^2$. }%
This large conversion at the synoptic scales 
and the spectral transfer of APE from the planetary scales 
toward the synoptic scales are mainly due to the baroclinic instability.
It is interesting to compare these results 
with the spectral energy budget of an equilibrated Eady flow 
\cite[]{MolemakerMcWilliams2010}.
The general picture is very similar with 
a dominance of the APE flux at large scales
and a stronger KE flux at small scales.
The increase of the total nonlinear flux at wavelengths between 700~km to 2000~km indicates that there is a direct forcing of APE at these scales,
most probably due to latent heat release organized at large scales.
This interpretation is consistent with results by \cite{Hamilton_etal2008}, 
who reported spectral magnitude at the mesoscales much smaller for the dry dynamical core than for the full AFES.

As already shown, 
the KE is mainly forced (by a conversion of APE) 
over a range of wavelengths between 1000~km to 5000~km.
At larger scales, the KE spectral flux is negative 
and reaches a minimum value approximately equal to -0.5~W/m$^2$.
A portion of the KE is transferred upscale, toward the planetary scales, 
and feeds the large-scale zonal winds.
At smaller scales, the KE spectral flux reaches a plateau at 0.62~W/m$^2$.
This shows that a non-negligible portion of the KE cascades toward small scales at the mesoscales. 
The value 0.62~W/m$^2$, which after conversion gives $6.1\times 10^{-5}$ m$^2$/s$^3$, 
is consistent with previous estimations 
for the KE spectral flux and for the small-scale dissipation rate \cite[]{ChoLindborg2001}.
Remarkably, 
the cumulative conversion $\CC[l]$ increases at the mesoscales 
(the local conversion is negative),
showing that the conversion is from KE to APE.
This demonstrates that the KE $k^{-5/3}$ spectrum is not produced by direct forcing of the KE.
Note that this conversion from KE to APE is consistent with strongly stratified turbulence.

% % % % One should not interpret the wavenumber where $\Pi_K[l]$ changes sign as the wavenumber 
% % % % where both upscale and downscale fluxes start.
% % % % These two processes actually coexist in the same range of scales.
The nonlinear KE spectral flux computed only with the rotational flow, $\Pi_{rot}$, 
is plotted as a red dotted line.
The interactions between the rotational modes conserve 
both KE and enstrophy ($|\bnablah \wedge \uu|^2/2$), 
exactly as in two-dimensional turbulence. 
Note that $\Pi_{rot}[l]$ is the spectral flux computed 
when the framework based on the two-dimensional vorticity equation is adopted, 
as for example in \cite{BoerShepherd1983}.
We see that these interactions are responsible for the upscale flux, 
which actually starts at wavelengths of the order of 2000~km.
In contrast, the complementary flux, $\Pi_K[l] - \Pi_{rot}[l]$, (red dashed line) 
is responsible for the downscale energy flux which starts at wavelengths of the order of 3000~km.
Indeed, the downscale energy cascade is produced by interactions involving the divergent part of the velocity field
\cite[]{MolemakerMcWilliamsCapet2010, DeusebioVallgrenLindborg2013}.

As shown by \cite{Lambert1984}, 
the variations of the cumulative conversion at very small wavenumber $l<8$ are due to the Hadley and Ferrel circulations.
The decrease of $\CC[l]$ at $l=2$ corresponding to 
a conversion of APE to KE of approximately 0.6~W/m$^2$ is mainly the signature of the Hadley cell.
The increase of $\CC[l]$ at $l=4$ corresponding to 
a conversion of KE to APE of approximately 0.5~W/m$^2$ is mainly the signature of the Ferrel cell.
The conversion at $l=2$ is weaker than previously computed by \cite{Lambert1984}.
Further investigations are necessary to understand 
if this effect is due to averaging over an insufficient amount of data or if it is a robust aspect of the AFES.

Figure \ref{fig_dyn_whole_atm}(b) presents the globally integrated spectral fluxes 
and cumulative conversion
as in figure \ref{fig_dyn_whole_atm}(a) but for the ECMWF model.
The planetary-scale features are overall quite similar to the results for the other model
even though the conversion at wavenumbers smaller than 4 
corresponding to the Hadley cell is much stronger
\Add{
($\Delta\CC \simeq 1$~W/m$^2$).
The total conversion $\CC[l=0]$ is equal to 2.9~W/m$^2$, 
consistent with results by \cite{BoerLambert2008} 
who computed averaging over a more longer time 
a total conversion for the ECMWF model of 3.1~W/m$^2$. }%
The baroclinic instability leads to large-scale positive APE flux 
and strong conversion at the synoptic scales 
\Add{$\Delta\CC \simeq 1.5$~W/m$^2$.}
However, the KE flux increases much less than for the AFES, 
indicating that there is strong dissipation at wavelengths of the order of 2000~km.
The total spectral flux for the AFES is also plotted in dashed black line for comparison.
The total flux for the ECMWF model is slightly smaller than for the AFES at the synoptic scales 
and is negative and very small at the mesoscales.
This is related to the weakness of the downscale mesoscale KE cascade ($\Pi_K[l]\simeq 0.15$~W/m$^2$) 
and to the fact that the APE flux is negative ($\Pi_A[l]\simeq -0.3$~W/m$^2$).
This unexpected result could be due to direct forcing of APE 
by release of latent heat at the smallest resolved scales of the order of 50~km.
This interpretation is consistent with the sign of the local conversion at the mesoscales, 
from APE to KE, in contrast to the case of the AFES.

\Add{
We shall now present a quantitative comparison between the two models.
In a stationary state, the amount of KE which is dissipated 
in a wavenumber range $[l_1,\ l_2]$ can be estimated as
\begin{equation}
\Delta\DD_K \simeq \Delta\CC + \Delta\Pi_K,
\label{eq_DeltaDDK}
\end{equation}
where
$\Delta\CC \equiv \CC[l_1]-\CC[l_2]$ is the amount of APE converted to KE in the range $[l_1,\ l_2]$ and
$\Delta\Pi_K \equiv \Pi_K[l_1]-\Pi_K[l_2]$ is the net amount of energy going into the range $[l_1,\ l_2]$ by the nonlinear fluxes at $l_1$ and $l_2$.
The effective forcing of APE can be evaluated in the corresponding way as
\begin{equation}
\Delta\GG - \Delta\DD_A \simeq \Delta\CC - \Delta\Pi_A.
\end{equation}
Finally, the effective total forcing can be evaluated as
\begin{equation}
\Delta\GG -\Delta\DD \simeq -\Delta\Pi,
\label{eq_DeltaGGmDD}
\end{equation}  
where $\DD[l] = \DD_K[l]+\DD_A[l]$ is the total energy dissipation 
and $\Pi[l]$ the total energy spectral flux.
Any increase (respectively decrease) of the total energy flux 
implies a net positive (respectively negative) forcing of the total energy.}

\begin{table}[H]
\caption{Quantitative energy budget for the range of wavenumbers 
going from $l_1= 12$ to $l_2 = 40$,
i.e., for wavelengths between 1000~km and 3200~km corresponding approximately to the synoptic scales,
for the AFES and the ECMWF models.
All values are in~W/m$^2$.
The quantities $\Delta\CC$, $\Delta\Pi_K$, $\Delta\Pi_A$ and $\Delta\Pi$ are directly obtained from figure~\ref{fig_dyn_whole_atm}.
The other quantities are evaluated using the equations (\ref{eq_DeltaDDK}-\ref{eq_DeltaGGmDD}).
}
\centerline{\begin{tabular}{ccc} 
                         & AFES  & ECMWF \\
 $\Delta\CC$     & ~0.73 & ~1.23 \\
 $\Delta\Pi_K$   & -0.63 & -0.51 \\
 $\Delta\Pi_A$   & ~0.54 & ~1.83 \\
 $\Delta\Pi$     & -0.09 & ~0.32 \\
                         &       &       \\
 $\Delta\DD_K$   & ~0.10 & ~0.72 \\
 $\Delta\GG-\Delta\DD_A$ 
                         & ~0.19 & ~0.41 \\
 $\Delta\GG-\Delta\DD$   
                         & ~0.09 & -0.31   
\end{tabular}}
\label{table_synoptic_scales}
\end{table}

\begin{table}[H]
\caption{\Add{Same as table~\ref{table_synoptic_scales} but for $l_2$ equal to the largest wavelength resolved by the model and for $l_1= 100$, which corresponds to a wavenumber of approximately 400~km.
This range of wavenumbers includes the mesoscales and the smallest length scales resolved by the models.
}}
\centerline{\begin{tabular}{ccc} 
                         & AFES  & ECMWF \\
 $\Delta\CC$     & -0.13 & ~0.11 \\
 $\Delta\Pi_K$   & ~0.62 & ~0.13 \\
 $\Delta\Pi_A$   & ~0.19 & -0.30 \\
 $\Delta\Pi$     & ~0.81 & -0.17 \\
                         &       &       \\
 $\Delta\DD_K$   & ~0.48 & ~0.24 \\
 $\Delta\GG-\Delta\DD_A$ 
                         & -0.32 & ~0.41 \\
 $\Delta\GG-\Delta\DD$   
                         & -0.81 & ~0.17   
\end{tabular}}
\label{table_smallest_scales}
\end{table}

\Add{
Table~\ref{table_synoptic_scales} summarizes the main differences between the two models 
in terms of the energy budget of the synoptic scales,
with $l_1 = 12$ and $l_2 = 40$,
corresponding to wavelengths between 1000~km and 3200~km.
The most important difference for these scales is the amount of kinetic energy dissipation,
$\Delta\DD_K$, 
which is equal to 0.10~W/m$^2$ for the AFES model and to 0.72~W/m$^2$ for the ECMWF model.
As already mentioned, the ECMWF model is very dissipative at the synoptic scales.
We have verified that a large part of this synoptic-scale dissipation 
takes place in the free atmosphere, i.e., far from the surface.
This unexpected and anomalous result 
could explain the lack of downscale energy cascade at the mesoscales observed for this model.
For both models, there is a net positive effective forcing of APE, $\Delta\GG-\Delta\DD_A$,
which is the signature of release of latent heat organized at the synoptic scales.
For the AFES, this release of latent heat is larger than the KE dissipation 
($\Delta\GG-\Delta\DD>0$),
which leads to an increase of the total energy flux $\Pi[l]$.
In contrast, for the ECMWF model the KE dissipation is larger than 
the effective forcing of APE ($\Delta\GG-\Delta\DD<0$)
so that the total flux decreases to approximately zero at $l=1000$.
}%

\Add{
Table~\ref{table_smallest_scales} presents a similar energy budget 
as table~\ref{table_synoptic_scales}
but for $l>100$,
i.e., for the mesoscales and the smallest length scales resolved by the models.
Only a small fraction (0.24~W/m$^2$) of the total KE dissipation ($\CC[0] = 2.9$~W/m$^2$) 
takes place at these scales for the ECMWF model
whereas these scales account for a significant part of the KE dissipation for the AFES.
Remarkably, the net APE forcing
$\Delta\GG-\Delta\DD_A$
is negative for the AFES (-0.32~W/m$^2$) and positive for the ECMWF model (0.41~W/m$^2$).
In this model, the APE and the KE are directly forced at small scales 
by latent heat release and conversion characterized by a weak spatial coherence.
}

\subsection{Vertically integrated non-dimensional spectra}

\begin{figure*}
\centering
% \centerline{
% \includegraphics[width=22pc]{Fig/fig_spectra_atm_AFES_T639.eps}
% \includegraphics[width=22pc]{Fig/fig_spectra_atm_ECMWF_T1279winter.eps}
% }
\centerline{\includegraphics[width=42pc]{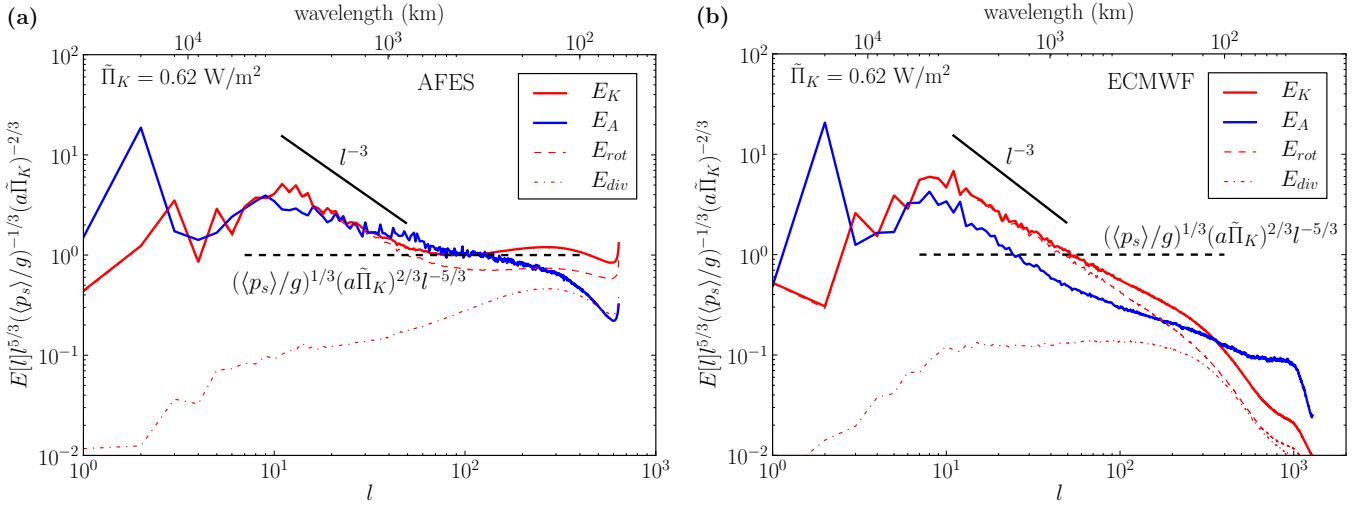}}
\caption{
Non-dimensional compensated spectra versus total wavenumber
for 
(a) the AFES T639 simulation and 
(b) the ECMWF IFS T1279 simulation.
The black dashed line represents the prediction (\ref{eq_prediction_spectra})
with $\tilde\Pi_K = 0.62$
(value marked by a cross in figure~\ref{fig_dyn_whole_atm}a).
The continuous straight line indicates the $l^{-3}$ power law.
}
\label{fig_spectra_whole_atm}
\end{figure*}

If it is assumed that the $l^{-5/3}$ range of the kinetic energy spectrum 
is of a similar form as the Kolmogorov spectrum of isotropic turbulence, 
then $E_K \propto \Pi_K^{2/3} l^{-5/3}$. 
If it is further assumed that the only other parameters that determine the spectrum are 
the radius of the Earth, $a$, 
and the total mass per unit area, which is equal to $\langle p_s \rangle/g$, 
then dimensional considerations give
\begin{equation}
E_K[l] = C  ( \langle p_s \rangle/g )^{1/3}  (a \Pi_K)^{2/3} l^{-5/3},
\label{eq_prediction_spectra}
\end{equation}
where $ C $ is a constant supposedly of the order of unity.
In the following, we choose as the typical KE flux the maximum of the KE flux: 
$\tilde\Pi_K = \max({\Pi_K}) = 0.62$.
(This value is marked by a cross in figure~\ref{fig_dyn_whole_atm}a.)
In figure \ref{fig_spectra_whole_atm}(a) the non-dimensional compensated spectra 
$ E[l] l^{5/3}  ( \langle p_s \rangle/g )^{-1/3}  (a \tilde\Pi_K)^{-2/3} $
for the AFES model
are plotted as a function of the total wavenumber.
At $l=1$, the APE spectrum (blue line) is much larger than the KE spectrum (red line),
as predicted by \cite{Lorenz1955}.
This leads to a ratio mean APE over mean KE approximately equal to 3.
However, for other wavenumbers (except at the largest ones), 
both spectra are of the same order of magnitude.
At the synoptic scales, the KE spectrum is quite steep.
It shallows at the mesoscales and presents a flat plateau corresponding to a $l^{-5/3}$ inertial range 
from 650~km up to the large-wavenumber dissipative range.
Remarkably, the constant $C$ in equation (\ref{eq_prediction_spectra}) is very close to unity,
which indicates 
that the $l^{-5/3}$ mesoscale range may be explained 
in a similar way as \cite{Kolmogorov1941} explained the $k^{-5/3}$ range of isotropic turbulence.
At wavelengths between 700~km to 2000~km, 
the APE spectrum (blue line) is equal to or larger than the KE spectrum 
and there are fluctuations resembling noise. 
However, the fluctuations do not seem related to lack of statistics.
They could be due to the direct forcing of APE at this scales (see figure \ref{fig_dyn_whole_atm}a).

In figure \ref{fig_spectra_whole_atm} are also plotted 
the rotational spectrum $E_{rot}[l]$ and the divergent spectrum $E_{div}[l]$
\Add{computed as in (\ref{eq_def_EKlm}).} %
At large scales, the rotational spectrum $E_{rot}[l]$ (dashed red line)
totally dominates over the divergent spectrum $E_{div}[l]$ (dashed dotted red line). 
Remarkably, the compensated divergence spectrum increases with $l$, 
meaning that $E_{div}[l]$ is shallower than a $l^{-5/3}$ power law.
Such very shallow divergent spectra were also obtained by simulations 
of strongly stratified and strongly rotating turbulence forced in geostrophic modes \cite[]{DeusebioVallgrenLindborg2013}
and of strongly stratified turbulence forced with rotational modes \cite[]{AugierBillantChomaz2013}.
However, other simulations with higher resolution would be necessary in order to check
whether the divergent spectra are sensitive to model parametrizations and resolution.
At the mesoscales, both spectra are of the same order of magnitude 
even though $E_{rot}[l]$ is larger than $E_{div}[l]$.

Figure \ref{fig_spectra_whole_atm}(b) shows the non-dimensional compensated spectra for the ECMWF model 
(we use the same value as for the AFES of $\tilde\Pi_K$ in order to allow an easier comparison between both models).
The planetary-scale features are quite similar to the AFES. 
At the synoptic scales, the ratio $E_K[l]/E_A[l]$ is approximately equal to 2,
indicating that the energy is partitioned equally between the two components of KE and the APE, 
as predicted by \cite{Charney1971}.
In contrast to the KE spectrum,
the APE spectrum becomes more shallow at high wavenumbers.
This is probably due to direct forcing of APE at the mesoscales.
Interestingly, the compensated divergent spectrum is flat, which means that it follows a $l^{-5/3}$ power law.
However, its magnitude is very small so that the vertically integrated KE spectrum 
at the mesoscales is nearly not affected.
It is interesting to note the similarity with 
the $k^{-5/3}$ divergent spectra obtained by \cite{WaiteSnyder2009} 
simulating a baroclinic life cycle.

\subsection{Vertical decomposition and vertical energy fluxes}

\begin{figure*}
\centering
% \centerline{
% \includegraphics[width=21pc]{Fig/fig_dyn_UPPERTROPO_AFES_T639.eps}
% \includegraphics[width=21pc]{Fig/fig_dyn_UPPERTROPO_ECMWF_T1279winter.eps}
% }
% \centerline{
% \includegraphics[width=21pc]{Fig/fig_dyn_STRATO_AFES_T639.eps}
% \includegraphics[width=21pc]{Fig/fig_dyn_STRATO_ECMWF_T1279winter.eps}
% }
\centerline{\includegraphics[width=42pc]{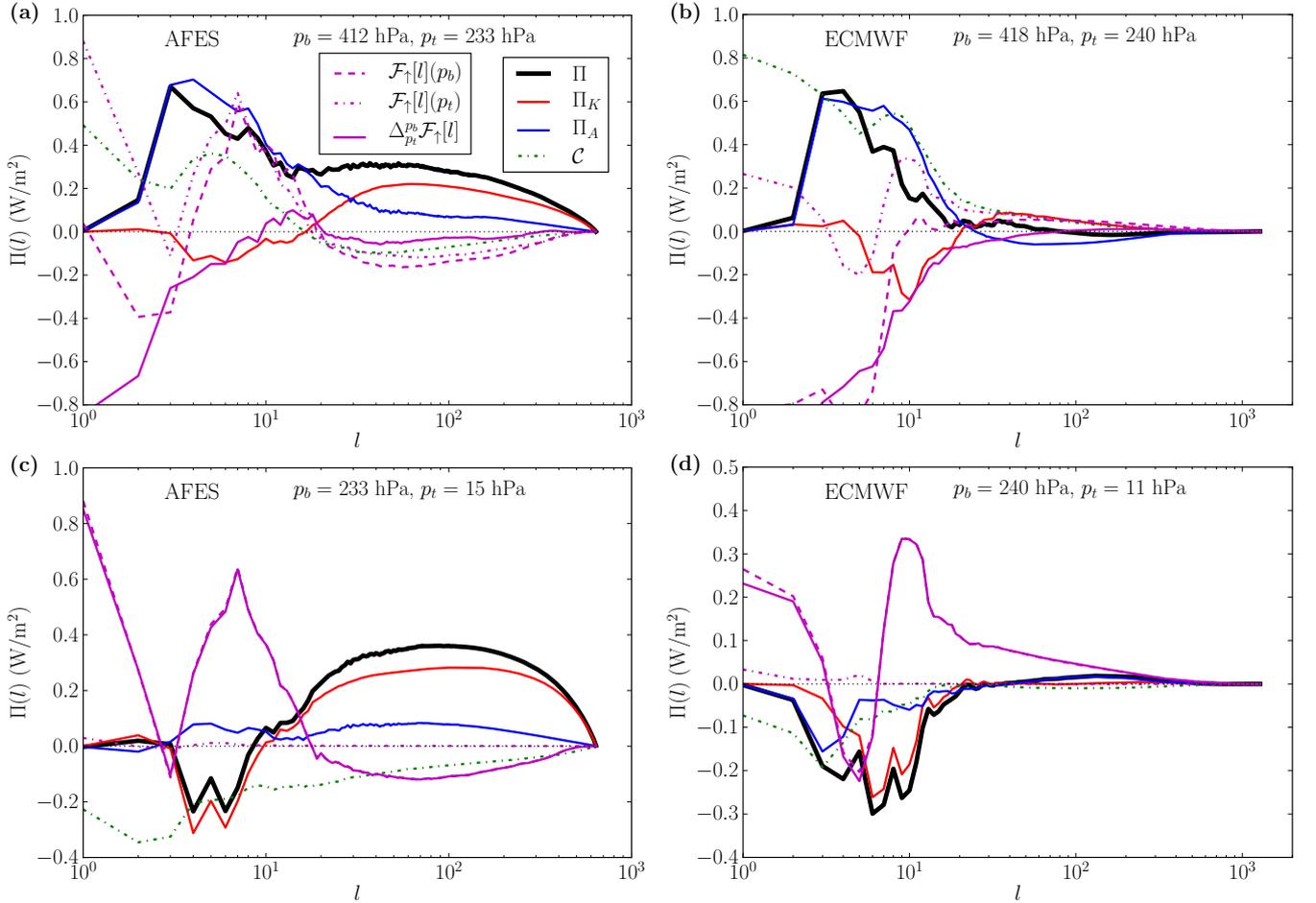}}
\caption{Same as figures \ref{fig_dyn_whole_atm} but integrated over layers corresponding approximately to 
(a,b) the stratosphere
and 
(c,d) the upper troposphere.
Figures (a,c) correspond to the AFES T639 simulation and 
figures (b,d) to the ECMWF IFS T1279 simulation.
The legend is given in (a) and $\Delta_{p_t}^{p_b}\mathcal{F}_{\uparrow}[l]= \mathcal{F}_{\uparrow}[l](p_b) - \mathcal{F}_{\uparrow}[l](p_t)$ is the cumulative inward vertical flux.
}
\label{fig_dyn_layers}
\end{figure*}

Figure~\ref{fig_dyn_layers} presents the spectral fluxes and the cumulative conversion
integrated over two different layers corresponding approximately to 
the upper troposphere (a,b)
and 
the stratosphere (c,d).
Figures (a,c) correspond to the AFES T639 simulation and 
figures (b,d) to the ECMWF IFS T1279 simulation.
Figure~\ref{fig_dyn_layers} also presents
the cumulative total vertical fluxes, $\mathcal{F}_{\uparrow}[l](p_b)$, at the bottom, 
and $\mathcal{F}_{\uparrow}[l](p_t)$, at the top of the layer
(magenta dashed and dashed-dotted lines, respectively).
The balance between these two inward and outward terms,
$\Delta_{p_t}^{p_b}\mathcal{F}_{\uparrow}[l]= \mathcal{F}_{\uparrow}[l](p_b) - \mathcal{F}_{\uparrow}[l](p_t)$,
is plotted in magenta as a continuous line.

For the AFES simulation, 
the spectral fluxes and the cumulative conversion 
integrated over the upper troposphere
(figure~\ref{fig_dyn_layers}a)
present roughly the same features as 
the globally integrated terms shown in figure~\ref{fig_dyn_whole_atm}(a).
The KE flux in the upper troposphere accounts for approximately 
one third of the globally integrated KE spectral flux.
The cumulative vertical flux $\Delta_{p_t}^{p_b}\mathcal{F}_{\uparrow}[l]$
is relatively small at the mesoscales and at the synoptic scales,
indicating that at leading order, the energy budget in this layer 
and at these scales is dominated 
by the spectral fluxes rather than by the vertical fluxes.
However, at the top and the bottom of the layer,
there are large vertical fluxes which are approximately equal.
These fluxes through the upper troposphere 
(upward
at wavenumbers $7\leqslant l\leqslant 20$ 
and downward
at wavenumbers $3 \leqslant l\leqslant 6$) 
account for exchanges of energy between the lower troposphere and the stratosphere.
At wavenumbers $1\leqslant l\leqslant 2$, $\mathcal{F}_{\uparrow}[l](p_t)$ decreases from 0.8~W/m$^2$ to 0 ~W/m$^2$
indicating a strong upward vertical flux at $p_t=233$~hPa.
Since the vertical flux at the bottom layer is small at these wavenumbers,
the layer loses energy.
The wavenumber ranges of the upward and downward fluxes at the planetary scales
nearly coincide with the wavenumber ranges of the conversion due to the Hadley and Ferrel cells, 
which seems to indicate that these vertical fluxes are related to the large-scale cells.
In contrast, the upward flux at the synoptic scales
indicates the presence of upward propagating planetary waves.
At $l>70$, $\mathcal{F}_{\uparrow}[l](p_b)$ and $\mathcal{F}_{\uparrow}[l](p_t)$ increase meaning 
that there is a downward vertical flux at these scales.
This shows that in the AFES, 
the mesoscales of the upper troposphere are not directly forced by upward propagating gravity waves,
as was the case in the simulation by \cite{KoshykHamilton2001} using the SKYHI model.

Figure~\ref{fig_dyn_layers}(b)
presents the same quantities as figure~\ref{fig_dyn_layers}(a),
also integrated over the upper troposphere
but for the ECMWF model.
Comparing figures~\ref{fig_dyn_layers}(a) and \ref{fig_dyn_layers}(b), 
we see the same differences 
as we saw between figures~\ref{fig_dyn_whole_atm}(a) and \ref{fig_dyn_whole_atm}(b).
The vertical fluxes are also quite different from the AFES
with small upward fluxes at the mesoscales 
and a smaller magnitude of the variations of the cumulative fluxes at large scales.

Figure~\ref{fig_dyn_layers}(c) shows the same quantities 
as figure~\ref{fig_dyn_layers}(a), also for the AFES
but integrated over the stratosphere.
At the large scales, 
the spectral fluxes and the cumulative conversion 
are quite different from the terms integrated over the height of the atmosphere and over the upper troposphere.
The APE flux has no large peak at $l\simeq 4$ and there is a conversion from KE to APE at synoptic scales 
rather than the other way around, as in the upper troposphere.
On the other hand, there is a strong upward energy flux at large scales from the bottom layer at $p_b = 233$~hPa 
while the corresponding flux at the top layer at $p_t = 15$~hPa is very small.
Thus, the stratosphere is not directly forced by baroclinic instability 
but by an energy flux from the troposphere, 
due to upwards propagating planetary waves 
and possibly also the effects of the Hadley and the Ferrel cells (see figure~\ref{fig_dyn_layers}a).
At the mesoscales, 
there is a conversion of KE into APE and a strong downscale cascade of KE and APE 
accounting for approximately half the globally integrated flux.
At $l>80$, $\Delta_{p_t}^{p_b}\mathcal{F}_{\uparrow}[l] \simeq \mathcal{F}_{\uparrow}[l](p_b)$ increases, 
meaning that the stratosphere loses energy by a downward flux through the tropopause.

Figure~\ref{fig_dyn_layers}(d) 
presents the same quantities as figure~\ref{fig_dyn_layers}(c),
also integrated over the stratosphere
but for the ECMWF model.
Note that the vertical axis is different from figure~\ref{fig_dyn_layers}(c).
In contrast to the AFES, 
\Add{the downscale energy cascade is very small}
% % % % there is nearly no downscale energy cascade 
and the stratospheric mesoscales are directly forced by an upward energy flux.

\begin{figure}[H]
\centering
% \centerline{\includegraphics[width=21pc]{Fig/fig_spectra_layers_AFES_T639.eps}}
% \centerline{\includegraphics[width=21pc]{Fig/fig_spectra_layers_ECMWF_T1279winter.eps}}
\centerline{\includegraphics[width=22pc]{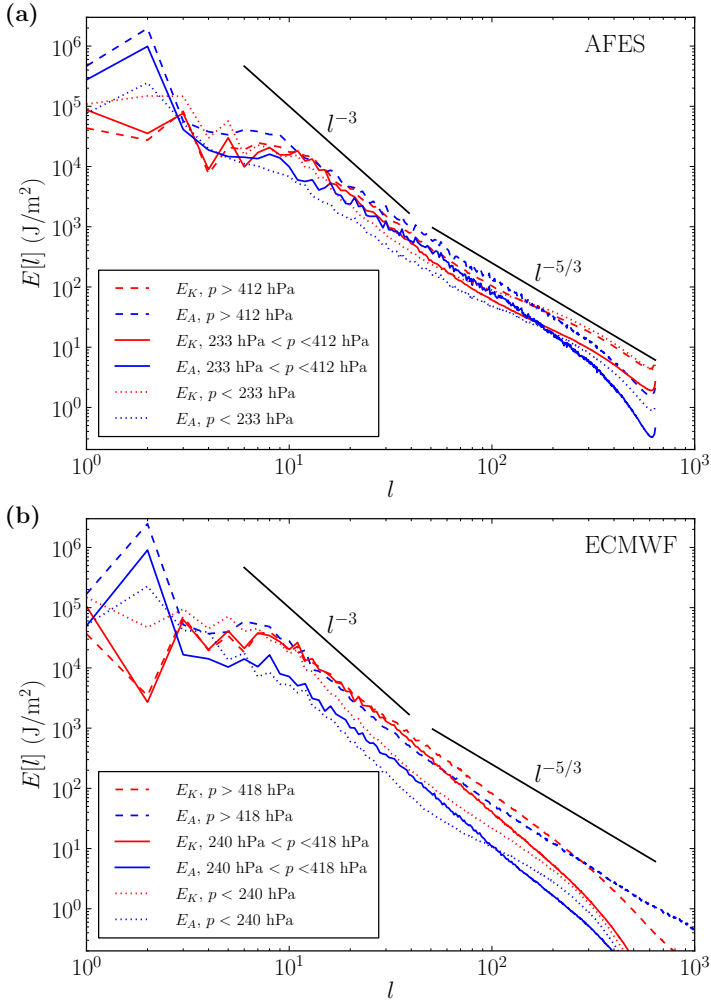}}
\caption{
KE and APE spectra integrated over three layers corresponding approximately to 
the lower troposphere,
the upper troposphere
and
the upper atmosphere.
The straight black lines indicate the $l^{-3}$ and $l^{-5/3}$ dependencies 
as guides for the eye.
Note that the axes and the straight lines are exactly the same in (a) and (b). 
}
\label{fig_spectra_layers}
\end{figure}

Figure \ref{fig_spectra_layers}
presents the KE and APE spectra integrated over 
three layers corresponding approximately to 
the lower troposphere,
the upper troposphere
and
the upper atmosphere,
for the AFES (figure \ref{fig_spectra_layers}a) 
and the ECMWF model (figure \ref{fig_spectra_layers}b).
Note that the axes and the straight lines are exactly the same in both figures.
At the large scales the spectra for both models are similar 
but 
at the mesoscales the spectral magnitude of the ECMWF is much smaller than the magnitude of the AFES.
Spectra integrated over the different layers lie closer to each other for the AFES than for the ECMWF.
For the AFES the KE spectra have similar shapes at different layers, whereas for the ECMWF model, 
they are quite different 
with a shallowing at the mesoscales only in the stratosphere (blue and red dotted lines)
as reported by \cite{BurgessErlerShepherd2013}.
However, these shallow spectra are 
much smaller in magnitude than the corresponding spectra from the AFES model
(more than one order of magnitude smaller at $l\simeq 100$).
From figure~\ref{fig_dyn_layers}(d) it is quite clear 
that the shallowing of the stratospheric spectra of the ECMWF model is not caused by an energy cascade 
but rather by gravity waves propagating from the troposphere.
The integrated spectra of the ECMWF model (figure ~\ref{fig_spectra_whole_atm}b) 
is dominated by the contribution from the troposphere.
In the lower troposphere, the APE spectrum presents a $l^{-5/3}$ dependency 
and is larger than the KE spectrum at large wavenumbers.
However, this cannot be explained by a downscale energy cascade
since the APE spectral flux is small and negative.
These differences between the APE and the KE spectra could be due to a damping of the KE 
and to a direct diabatic forcing of the APE at the very large wavenumbers $l \sim 1000$.
For the AFES model (figure \ref{fig_spectra_layers}a), the spiky irregular shape
observed at the synoptic scales for the globally integrated APE spectrum (figure \ref{fig_spectra_whole_atm}a)
is also observed for the tropospheric APE spectra 
but not for the stratospheric APE spectrum.
This could indicate that this effect is related to the topography.
However, the spikes are not only present in the lower but also in the upper tropospheric spectra
whereas the upper troposphere is not pierced by the topography.
Moreover, the vertical fluxes at $p = 412$~hPa are rather regular.
Therefore, the spiky irregular shape of the APE spectra seems to be due to processes
happening in the troposphere and not only at the surface.
The release of latent heat organized at the synoptic scales could be an explanation.

\section{Conclusions}

A new formulation of the spectral energy budget has been presented 
and applied to study the results of two different GCMs.
In contrast to previous formulations,
both KE and APE are considered 
and the topography is taken into account.
Moreover,
the advection terms are exactly separated into spectral transfer and vertical flux 
and the pressure term is exactly separated 
into adiabatic conversion and vertical flux.

The spectral fluxes show that the AFES, 
which 
% % % % reproduces realistic spectra with a $k^{-5/3}$ power law at the mesoscales, 
\Add{produces realistic $k^{-5/3}$ KE spectra at the mesoscales,}
simulates a strong downscale energy cascade of $\Pi[l] \simeq 0.8$~W/m$^2$. 
The vertical fluxes for the upper troposphere show that
the mesoscales are not directly energized by gravity waves propagating from the ground.
Moreover, the spectra collapse on the prediction based on the existence of the cascade,
indicating that the mesoscale $k^{-5/3}$ power law is due to the downscale energy cascade. 
In contrast, neither the $k^{-5/3}$ \Add{KE spectrum }% 
nor the downscale energy cascade are produced by the ECMWF model.
%
% % % % The study of the spectra and their tendencies integrated over different layers reveals that 
% % % % in the AFES, the stratospheric mesoscales are not forced by upward propagating gravity waves. 
% % % % In contrast to this, the stratospheric spectra of the ECMWF model are directly forced at the mesoscales 
% % % % by gravity waves propagating from the troposphere.
\Add{
The analysis of the spectra and their tendencies integrated over different layers reveals that in the ECMWF, the stratospheric mesoscales are directly forced 
by gravity waves propagating from the troposphere.
In contrast to this, the stratospheric spectra of the AFES are forced by the downscale energy cascade and at the mesoscales, gravity waves produced in the stratosphere propagate to the troposphere.}

These results show that our spectral energy budget formulation is a convenient tool
to investigate the issue of the mesoscale dynamics and its simulation by GCMs.
In particular, 
we have shown that
the flux computed only with the rotational part of the velocity field $\Pi_{rot}$ 
\cite[as i.e., in][]{Fjortoft1953,BoerShepherd1983}
accounts only for the upscale KE flux.
Since the downscale energy cascade is produced 
by interactions involving the divergent part of the velocity field
\cite[]{VallgrenDeusebioLindborg2011, DeusebioVallgrenLindborg2013, MolemakerMcWilliamsCapet2010},
one must consider 
the nonlinear transfers computed with the total advection term.
To do this in a consistent way, 
the advection terms in the spectral energy budget have to be split 
as shown in section~\ref{section_formulation}.

\Add{
Our results raise many interesting questions.
In particular,
we have shown that the two studied models simulate very different dynamics at the mesoscales.
The AFES model reproduces a downscale energy cascade at the mesoscales,
whereas such a cascade is absent in the ECMWF model.
One important difference between the two models is 
the vertical resolution.
The vertical grid space in the high troposphere in approximately 
500 m for the ECMWF model and 2~km for the AFES model.
A coarser resolution should be a drawback for simulating the mesoscale dynamics,
both gravity waves and strongly stratified turbulence.
The ECMWF model can simulate gravity waves with smaller vertical resolution 
than those simulated by the AFES model. 
Stratified turbulence is very demanding in terms of vertical resolution, 
since it is necessary to resolve the buoyancy length scale, $L_b \sim U/N$, 
which is the characteristic vertical length scale of horizontal layers 
\cite[]{WaiteBartello2004, Lindborg2006, Waite2011}. 
Here, $U$ is the characteristic velocity of the horizontal wind 
and $N$ is the Brunt-V\"ais\"al\"a frequency.  
Since $L_b$ is of the order of 1~km in the atmosphere, 
none of the models really resolve this length scale. 
The question is then, how is it possible that any of these two models is able 
to simulate a downscale energy cascade? 
Our tentative answer to this question is that vertical resolution does not seem to be very critical 
in order to reproduce a downscale energy cascade, 
because such a cascade seems to be a very general property 
of rotating and stratified hydrodynamic systems 
which are not too close to the quasigeostrophic regime. 
That a downscale energy cascade emerges in such systems has been demonstrated 
in recent simulations \cite[]{MolemakerMcWilliamsCapet2010} 
that do not have extremely fine vertical resolution.
Dynamically, the downscale cascade is characterized by the importance of 
the interactions involving ageostrophic modes.
This raises the question of how ageostrophic motions at scales of the order of thousands of kilometers are produced from geostrophic motions at large scales. In particular, the baroclinic instability does not produce ageostrophic motions. Our results indicate that other ageostrophic instabilities 
%%%%\cite[]{McWilliamsYavnehCullenGent1998} 
are active and very important in the AFES model.}

\Add{
In subsection~\ref{section_application}\ref{subsection_vert_int_SEB}, we have discussed the energy budget of the synoptic scales
(see table~\ref{table_synoptic_scales}).
We have shown that the ECMWF model is very dissipative, even at these very large scales.
In the atmosphere, the dissipation takes place at scales of the order of one centimeter or smaller.
In a GCM, dissipation, namely loss of energy at resolved scales, 
has to correspond to physical processes transferring energy to unresolved scales.
It seems unlikely that physical processes could account for such large highly non-local transfer
from the synoptic scales to the unresolved scales,
which are relatively small for the ECMWF T1279L91 model.
This indicates that the synoptic-scale dissipation for the ECMWF model
is too large. 
Since energy is removed at the synoptic scales, it is not available 
to feed the downscale energy cascade.
Our interpretation of the data analysis is that the ECMWF model does not simulate the downscale cascade 
at the mesoscales
because of an excessive dissipation at the synoptic scales.
If this interpretation is correct, 
a decrease of the dissipation at the synoptic scales would have strong consequences 
in terms of the mesoscale dynamics.
Thus, it would be important to understand why the ECMWF model is so dissipative at the synoptic scales.
One explanation could be
numerical dissipation related to the semi-Lagrangian semi-implicit scheme.
However,
it seems unlikely that the numerical scheme could account for such large dissipation at scales
approximately 200 times larger than the typical horizontal grid scale, which is equal to 16~km.
Therefore, 
it would be interesting to explicitly compute 
the dissipation spectrum of all non-conservative terms related to e.g.
the turbulent scheme and the wave drag.}

Future investigations could also compare different models varying the resolution, 
the convection schemes, the advection scheme and the turbulent models.
It would be desirable to analyze simulations over long periods
to obtain better statistical convergence at large scales.
It could be informative to analyze idealized simulations 
using dry dynamical core and/or aquaplanet versions of the models
with variation of physical parameters such as the rotation rate and the large-scale forcing.
Other interesting aspects for future work are the consideration of the effects of the water content
and the adaptation of the formulation for non-hydrostatic simulations.

% \begin{acknowledgment} 
We wish to thank Kevin Hamilton and Nils Wedi for providing the data.
We thank Kevin Hamilton for showing nice hospitality in Hawaii,
Erland K\"all\'en and Nils Wedi for inviting us to ECMWF where we had interesting discussions 
and
Nathanael Schaeffer for providing his library for spherical harmonic transform SHTns 
\cite[]{Schaeffer2013}.

% \end{acknowledgment}

\appendix
\section*{Appendix A\ \ Technical details on topography in $p$-coordinates}

\label{section_appendixA}

Following \cite{Boer1982}, the equations are extended over a domain including subterranean pressure levels.
For any function $f(\xxh, p)$ whose values below the surface have been obtained by interpolation, 
we define a corresponding function $\tilde f(\xxh, p) =\beta(\xxh, p)f(\xxh, p)$,
where $\beta(\xxh, p) =  \mbox{H}(p_s(\xxh)-p)$ and $\mbox{H}$ is the Heaviside function.
In the extended domain, the boundary condition at the earth surface is expressed as $\D_t \beta = 0$,
with $\D_t = \p_t + \vv\cdot\bnabla$. 
Using the chain rules
$\p_t \beta = \delta \p_t p_s$, $\bnablah \beta = \delta \bnablah p_s$ and $\p_p \beta = -\delta$,
we recover the classical boundary condition for the atmospheric domain in pressure coordinates:
$ \D_t \beta = \delta ( \p_t p_s + \uu \cdot \bnablah p_s - \omega_s) = 0$,
where $\delta = \delta(p_s-p)$ is the Dirac distribution with impulse at the surface.
\Add{
Note that taking into account the topography has of course no influence for the high troposphere and the stratosphere.
However, the increase of the accuracy is important for the lower troposphere, 
especially through the computation of the representative mean temperature 
without using interpolated subterranean data.
}

\Add{
For each pressure level, even those pierced by the topography, 
the spectral energy functions are defined as
\begin{equation}
E_K^{[lm]}(p) = \frac{\scalarprodvec{\tuu}{\tuu}}{2}
\mbox{ \ and \ }
E_A^{[lm]}(p) = \gamma(p) \frac{|\tthetap_{lm}(p)|^2}{2}. \nonumber
\end{equation}
The spectral energy budget is derived as in section \ref{section_formulation}, 
i.e., using the relations
$\p_t E_K^{[lm]}(p) = \scalarprodvec{\tuu}{\p_t \tuu} $ and 
$\p_t E_A^{[lm]}(p) = \gamma(p) \scalarprod{\tthetap}{\p_t \tthetap}$.
For levels pierced by the topography, 
a surface term arises from the pressure term $\scalarprodvec{\tuu}{\beta \bnablah \Phi}$
and can be expressed as
\begin{align}
S^{[lm]}(p) = 
& - \scalarprod{\tilde\omega}{\delta \Phi_s} 
  - \scalarprod{\delta \p_t p_s}{\tilde\Phi} \nonumber\\
& + \scalarprodvec{\tuu}{\delta \Phi_s \bnablah p_s }. 
\end{align}
The sum over all spherical harmonics of $S^{[lm]}(p)$ 
is equal to  the surface term $S(p)$ in equation (\ref{eq_EKp}).
However, the term $S^{[lm]}(p)$ is difficult to evaluate and has not been computed.}

\Add{
The expressions of the other terms are very similar as those derived in section \ref{section_formulation}.
The nonlinear transfers can be written as
\begin{align}
T_K^{[lm]}(p)   = & -\scalarprodvec{\tuu}{\uu\cdot\bnablah \tuu + d \tuu/2} \nonumber \\
                  & +\big(   \scalarprodvec{\p_p\tuu}{\omega \tuu} 
                    - \scalarprodvec{\tuu}{\omega \p_p\tuu}  \big)/2,  \\
T_A^{[lm]}(p)   = & -\gamma       \scalarprod{\tthetap}{\uu \cdot \bnablah \tthetap + d \tthetap/2}  \nonumber\\
                  & +\gamma \big( \scalarprod{\p_p\tthetap}{\omega\tthetap} 
                                  - \scalarprod{\tthetap}{\omega\p_p\tthetap} \big)/2.
\end{align}
The linear transfer arising from the Coriolis strength is 
\begin{align}
L^{[lm]}(p) = f_0 \big( &    \scalarprod{\breve\psi}{\sin\varphi  \divh(\tuu) + \cos\varphi \p_\varphi \breve\chi} \nonumber\\
            & \hspace{0mm} + \scalarprod{\breve\chi}{\sin\varphi  \roth(\tuu) - \cos\varphi \p_\varphi \breve\psi}  \big).
\end{align}
where $\breve\psi$ and $\breve\chi$ are the stream function and velocity potential computed from $\tuu$.
The diabatic term, the conversion term and the diffusion terms are 
\begin{eqnarray}
G^{[lm]}(p)  & = &  \gamma  \scalarprod{\tthetap}{\tilde Q_{\theta}'},\\
C^{[lm]}(p)  & = & -        \scalarprod{\tilde\omega}{\tilde\alpha}, \\
D_K^{[lm]}(p)& = & -        \scalarprodvec{\tuu}{\beta \D_\uu(\uu)}.\\
D_A^{[lm]}(p)& = & - \gamma \scalarprod{\tthetap}{\beta\D_\theta(\theta)}.
\end{eqnarray}
The vertical fluxes and the non-conservative term can be computed as
\begin{eqnarray}
F_{K \uparrow}^{[lm]}(p) & = & -  \scalarprod{\tilde\omega}{\tilde\Phi} 
- \scalarprodvec{\tuu}{\omega \tuu}/2,\\
F_{A \uparrow}^{[lm]}(p) & = & - \gamma      \scalarprod{\tthetap}{\omega \tthetap}/2 , \\
J^{[lm]}(p)              & = & -(\p_p \log\gamma) F_{A \uparrow}^{[lm]}(p) .
\end{eqnarray}
}

In practice, computing the spherical harmonics transform of $\tuu = \beta \uu$ is not numerically feasible 
at pressure levels pierced by the topography, i.e., where $\beta$ is equal to 1 or to 0.
We must use a modified smooth $\beta$.
Moreover, the spherical harmonic transform of $\tuu = \beta \uu$ should not
reflect the spectral content of the topography rather than that of $\uu$.
Therefore, it is convenient to use an alternative $\beta$ computed with a time-averaged pressure surface 
$\overline{p_s}(\xxh)$ \cite[]{Boer1982}.
A smooth low-pass filter with a cutoff total wavenumber equal to 40 is then applied to this function 
$\mbox{H}( \overline{p_s}(\xxh)-p)$.
As a consequence, 
a large part of the interpolated subterranean data 
(mainly under the large topographic highs, i.e., Antarctic continent, Himalaya and Andean mountain ranges) 
are not used in the calculations, 
but the spectral quantities at relatively high wavenumbers are not affected by the high wavenumber content of the topography.

\section*{Appendix B\ \ Total energy conservation and APE}

\label{section_appendixB}

\Add{
In this appendix, we investigate the meaning of the surface term appearing in equation \ref{eq_EKp}
and
show how equations (\ref{eq_EKp}-\ref{eq_EAp}) can be related to the total energy conservation.
Neglecting all diabatic processes, 
it is straightforward to derive the local conservation equation
\begin{equation}
\D_t(E_K + H) = - \bnabla \cdot (\vv \Phi),
\end{equation}
with $E_K(\xxh, p) = |\uu|^2/2$.
By computing $\D_t(  \beta (E_K + H) ) $, we obtain an equation which is valid 
over a domain including subterranean pressure levels
\begin{equation}
\D_t(\tilde E_K + \tilde H) = - \beta \bnabla \cdot (\vv \Phi) = -\bnabla \cdot (\vv \tilde\Phi) + \Phi\vv\cdot\bnabla\beta.
\end{equation}
Taking the horizontal mean and integrating over pressure gives
\begin{equation}
\p_t \int_0^{\infty} \langle \tilde E_K + \tilde H \rangle dp/g = - \p_t \langle \Phi_s p_s \rangle/g,
\label{eq_EKplusH}
\end{equation}
where we have used the relations $\Phi\vv\cdot\bnabla\beta = -\Phi \p_t\beta = - \delta \Phi_s \p_t p_s  $
and the fact that $\Phi_s(\xxh)$ is not a function of time.
}

\Add{
Using the hydrostatic equation, 
it can be shown that $E_I + \Phi = H + \p_p (p \Phi)$,
which after integration gives
\begin{equation}
\p_t\langle \Phi_s p_s \rangle/g = \p_t \int_0^{\infty} \langle \tilde E_I + \tilde \Phi - \tilde H \rangle dp/g,
\label{eq_B4}
\end{equation}
where $E_I$ is the internal energy per unit mass.
Substituting this result into (\ref{eq_EKplusH}) finally gives the total energy conservation equation
\begin{equation}
\p_t \int_0^{\infty} \langle \tilde E_K + \tilde E_I + \tilde \Phi \rangle dp/g = 0.
\end{equation}
Note that (\ref{eq_EKplusH}) and (\ref{eq_B4}) can be rewritten as
\begin{align}
\p_t \int_0^{\infty} \langle \tilde E_K \rangle dp/g =
& \int_0^{\infty} C(p) dp/g   - \p_t\langle \Phi_s p_s \rangle/g  , \\
\p_t \int_0^{\infty} \langle \tilde E_I + \tilde \Phi \rangle dp/g = 
& -\int_0^{\infty} C(p) dp/g   + \p_t\langle \Phi_s p_s \rangle/g  , 
\end{align}
which shows that the term $\p_t\langle \Phi_s p_s \rangle/g$ 
can be interpreted as a conversion of $E_K$ 
into $E_I + \Phi$.
Using the hydrostatic equation we find that
$\langle \Phi_s p_s \rangle/g = \langle \Phi_s \rho_A \rangle$, 
where $\rho_A(\xxh, t)$
is the mass per unit area of the total depth of the atmosphere.
Thus, $\langle \Phi_s p_s \rangle/g$ may be interpreted as 
the mean potential energy per unit area of an
atmosphere where the centre of mass of each air column has been moved to the ground.
This quantity can change only if the density distribution is changed with respect to
topography, so that high or low density air on average is moved to high or low topography
regions.
}

\Add{
The sum of equations (\ref{eq_EKp}) and (\ref{eq_EAp}) integrated over pressure is
\begin{align}
\p_t \int_0^{\infty} \langle \tilde E_K + \tilde E_A \rangle dp/g 
= &\int_0^{\infty} \big(S(p) +J(p)\big) dp/g, \nonumber \\
= & - \p_t\langle \Phi_s p_s \rangle/g + \int_0^{\infty} J(p) dp/g.
\end{align}
The sum of the integrated KE and the Lorentz APE is not exactly conserved
and can be produced by two adiabatic non-conservative terms, 
the volumetric term $J(p)$ related to the non-linearity of the potential temperature profile
and the surface term $S(p)$.
%
% Therefore, the APE does not account for the conversion of $E_I + \Phi$ 
% into $E_K$ at the surface.
%
The total mass conservation can be written as $\p_t\langle p_s \rangle =0$,
which implies that $\p_t\langle \Phi_s p_s \rangle = \p_t\langle \Phi_s p_s' \rangle$,
where $p_s'= p_s-\langle p_s \rangle$.
Moreover, since $\Phi_s$ does not vary with time, we also have
$\p_t\langle \Phi_s p_s' \rangle = \p_t\langle \Phi_s (p_s'-\overline{p_s'}) \rangle$,
where the bar denotes the temporal average.
We have computed the quantity $\langle \Phi_s (p_s'-\overline{p_s'}) \rangle/g$ 
and it is very small compared 
to the horizontally averaged vertically integrated KE and APE.
This explains why it is relevant to neglect the topography as done by \cite{Lorenz1955}.
}

\bibliographystyle{ametsoc}

\bibliography{bib_article_AL_JAS}

\end{multicols}

\end{document}